\documentclass[%
superscriptaddress,
unsortedaddress,
runinaddress,
frontmatterverbose,
showpacs,
 amsmath,amssymb,
 aps,
 prl,
 lengthcheck,%
]{revtex4-1}
\usepackage{graphicx}
\usepackage{dcolumn}
\usepackage{bm}
\usepackage{hyperref}
\begin{document}

\title{
Expansion of Arbitrary Electromagnetic Fields in Terms of Vector Spherical Wave
Functions}
\thanks{
W. L. Moreira and C. L. Cesar belong to CEPOF-FAPESP and INFABIC-CNPq and are
grateful to FAPESP, CNPq, CAPES and PETROBRAS for financial assistance.
We also thank Rodrigo G. Pereira at Kavli Institute for Theoretical Physics
(UCSB) for theoretical discussions about the angular momentum operator in
k-space.
}
\author{W. L. Moreira     ${}^{1}$}
\author{A. A. R. Neves    ${}^{2}$}
\author{M. K. Garbos      ${}^{3}$}
\author{T. G. Euser       ${}^{3}$}
\author{P. St. J. Russell ${}^{3}$}\email{philip.russell@mpl.mpg.de}
\author{C. L. Cesar       ${}^{1}$}\email{lenz@ifi.unicamp.br}
\affiliation{${}^{1}$Instituto de F\'isica Gleb Wataghin, State University of
Campinas, 13.083-970 Campinas, S\~ao Paulo, Brazil}
\affiliation{\\${}^{2}$National Nanotechnology Laboratory, Istituto di
Nanoscienze of CNR, Universit\`a del Salento, via Arnesano, 73100, Lecce, Italy}
\affiliation{\\${}^{3}$Max Planck Institute for the Science of Light, 91058
Erlangen, Germany}

\date{\today}\begin{abstract}
Since 1908, when Mie reported analytical expressions for the fields scattered
by a spherical particle upon incidence of an electromagnetic plane-wave,
generalizing his analysis to the case of an arbitrary incident wave has proved
elusive. This is due to the presence of certain radially-dependent terms in the
equation for the beam-shape coefficients of the expansion of the
electromagnetic fields in terms of vector spherical wave functions. Here we
show for the first time how these terms can be canceled out, allowing
analytical expressions for the beam shape coefficients to be found for a
completely arbitrary incident field. We give several examples of how this
new method, which is well suited to numerical calculation, can be used.
Analytical expressions are found for Bessel beams and the modes of rectangular
and cylindrical metallic waveguides. The results are highly relevant for
speeding up calculation of the radiation forces acting on spherical
particles placed in an arbitrary electromagnetic field, such as in optical
tweezers.
\end{abstract}

\maketitle
Gustav Mie, in his celebrated 1908 paper \cite{Mie1908}, used the vector
spherical wave function (VSWF), or partial wave expansion (PWE), of a linear
polarized plane-wave to generalize scattering theories to spherical particles
of any size, from geometrical optics to the Rayleigh regime, and thus was able
to clarify many phenomena, for example in atmospheric physics. He obtained
analytical expressions for the expansion coefficients based on special
mathematical identities related to a plane-wave. This beam expansion was
necessary for applying boundary conditions at a spherical interface. Since
then, with the arrival of lasers and optical waveguides, the diversity and
complexity of possible incident fields has become enormous so that the
restriction to an incident plane-wave has become unrealistic.

Different experiments, ranging from particle levitation and trapping
\cite{Ashkin1970,Ashkin1987}, to the ultrahigh-Q microcavities used in
cavity QED experiments \cite{Collot1993,Vernooy1998}, use different beams. For
example, very high numerical aperture beams are used in optical tweezers and
confocal microscopy \cite{Svoboda1994,Hell1993,Brakenhoff1979}, evanescent
fields in near-field microscopy \cite{Betzig1992,Sanchez1999}, and the
waveguide modes of a fiber taper are employed to couple light to the
whispering gallery modes of spherical microcavities \cite{Cai2000}. Optical
forces, absorption, Raman scattering and fluorescence can be greatly enhanced
inside spherical microcavities at Mie resonances
\cite{Neves2006,Vollmer2002,Spillane2002,Ng2005}. Laguerre-Gaussian,
Hermite-Gaussian and Bessel beams \cite{Arlt2000,Novotny1998}, and the
internal electromagnetic field of hollow core photonic crystal fibers
\cite{Russell2003,Euser2009}, are used to trap and transport particles. The
understanding of all these phenomena requires a precise knowledge of the VSWF
coefficients of the incident beams. A generalized Lorenz-Mie theory was
developed to handle the many variants of beams beyond classical plane-waves,
and the expansion coefficients in these cases are known as beam shape
coefficients (BSC) \cite{Gouesbet1988,Gouesbet2009}. Moreover, because the
VSWFs form an orthogonal complete basis, they can be used to study scattering
and forces \cite{Nieminen2001} on non-spherical particles, and are the starting
 point of the powerful T-matrix methods \cite{Mishchenko1996}.

The calculation of BSCs for an arbitrary beam has always been a complicated
task, requiring significant effort. Furthermore, there is a fundamental problem
 with these calculations: an expansion of any function in some basis is
complete only when the expansion coefficients can be written in terms of scalar
 products, or integrals, with defined numerical values. This task has actually
never been accomplished for the VSWFs of an arbitrary beam because the integral
 over the solid angle does not explicitly eliminate the radial dependence, at
least up until now. So far as we are aware, the current literature lacks any
mathematical proof that this radial function, which appears after integration
over all solid angles for any type of beam that satisfies Maxwell's equations,
can exactly cancel out the spherical Bessel function that appears on the other
side of the BSC equation. If this is not true, then the BSC could not be a
constant independent of the radial coordinate -- as required for a successful
expansion.

This non-radial dependence of the BSC has been proven only for the case of
plane-waves and for a high numerical aperture focused Gaussian beam
\cite{Neves2006a}. Working with an electromagnetic mode inside a hollow
cylindrical waveguide, we also have been able to obtain analytical
expressions for constant BSCs that depend only on the position of the
reference frame. This raises the fundamental question, whether it would be
possible to prove that the spherical Bessel function would naturally emerge
from the solid angle integral for any type of electromagnetic field. The
purpose of this letter is to show, we believe for the first time, that this
is indeed possible. The implications of this result for computational light
scattering is very noteworthy. We show how the new method can be used to
calculate the BSCs for plane-waves, cylindrical and rectangular waveguide
modes and Bessel beams.

The dimensionless BSCs $G_{lm}^{TE/TM} $ for an incident field
$\bm{{\rm E}}=\bm{{\rm E}}(\bm{{\rm r}})$,
$\bm{{\rm H}}=\bm{{\rm H}}(\bm{{\rm r}})$ are defined in the
equations \cite{Jackson1999}
\begin{eqnarray}\label{eq:expansion}
\left[
\begin{array}{c}
\bm{{\rm E}}\\
Z\bm{{\rm H}}
\end{array}
\right]=E_0\!
\sum_{p,q}
\left[
\begin{array}{c}
G_{lm}^{TE}\\
G_{lm}^{TM}
\end{array}
\right]\!
\bm{{\rm M}}_{lm}
\!+\!
\left[
\begin{array}{c}
G_{lm}^{TM}\\
-G_{lm}^{TE}
\end{array}
\right]\!
\bm{{\rm N}}_{lm},
\end{eqnarray}
 $E_0 $ is an electric field dimension constant,
$k\bm{{\rm N}}_{lm}=i\nabla\times\bm{{\rm M}}_{lm}$,
$\bm{{\rm M}}_{lm}=j_l(kr)\bm{{\rm X}}_{lm}
(\bm{{\rm\hat{r}}})$, $j_l(kr)$ are spherical Bessel functions,
$\bm{{\rm X}}(\bm{{\rm\hat{r}}})=\bm{{\rm L}}Y_{lm}
(\bm{{\rm\hat{r}}})/\sqrt{l(l+1)}$ are the spherical harmonics,
$Z=\sqrt{\mu /\varepsilon }$, $k=\omega \sqrt{\mu \varepsilon }$ and
$\bm{{\rm L}}=-i \bm{{\rm r}}\times d/d\bm{{\rm r}}$, 
$d/d\bm{{\rm r}}=\bm{{\rm\hat{x}}}\partial_x+
\bm{{\rm\hat{y}}}\partial_y+\bm{{\rm\hat{z}}}\partial_z$ is the
gradient operator in the coordinates (direct) space.
Throughout this paper we use the convention that the terms inside [] are parts
of separate equations, i.e., (\ref{eq:expansion}) contains two equations,
the first relating $\bm{{\rm E}}$ to $G_{lm}^{TE} $ and $G_{lm}^{TM} $ and the
second, $Z\bm{{\rm H}}$ to $G_{lm}^{TM} $ and $-G_{lm}^{TE}$.

The usual procedure for obtaining the BSC's involves multiply both sides of
(\ref{eq:expansion}) by $\bm{{\rm X}}_{l'm'}^*$,
take scalar products with the fields and integrate over the solid angle
$\Omega$. Due to the orthogonality properties of the vector spherical
harmonics \cite{Jackson1999,Arfken2005}, one can easily show that
\begin{eqnarray}\label{eq:origbsc}
E_0j_l (kr)\!
\left[
\begin{array}{c}
G_{lm}^{TE}\\
G_{lm}^{TM}
\end{array}
\right]
\!&=&\!\int\! d\Omega(\bm{\hat{{\rm r}}})~\bm{{\rm X}}_{lm}^*(\bm{\hat{{\rm r}}})\!\cdot\!
\left[
\begin{array}{c}
\bm{{\rm E}}(\bm{{\rm r}})\\
Z\bm{{\rm H}}(\bm{{\rm r}})
\end{array}
\right]\!,
\end{eqnarray}
 $\Omega(\bm{\hat{{\rm r}}})$ is the solid angle with respect 
to an arbitray origin not related to any particular  point of the incident beam,
explicit said to be in the direct space. 
Equation
 (\ref{eq:origbsc}) does not yield explicit expressions for the BSCs because
the LHS still contains the radially-dependent spherical Bessel function. Our
goal is to extract this function from the RHS and cancel it out with the one
on the LHS, for any general incident electromagnetic field. To accomplish this
 we use the Fourier transform $\mathcal{F}$ of the fields
\begin{eqnarray}
\left[
\begin{array}{c}
\bm{{\rm E}}(\bm{{\rm r}})\\
\bm{{\rm H}}(\bm{{\rm r}})
\end{array}
\right]
&=&
\frac{1}{(2\pi )^{3/2}}\int  d^3k'
\left[
\begin{array}{c}
\bm{\mathcal{E}}(\bm{{\rm k}}')\\
\bm{\mathcal{H}}(\bm{{\rm k}}')
\end{array}
\right]
e^{i\bm{{\rm k}}'\cdot\bm{{\rm r}}}.
\end{eqnarray}
By this definition, one can show that
$\mathcal{F}\{\bm{{\rm L}}\psi(\bm{{\rm r}})\}
=\bm{{\mathcal{L}}}\Psi(\bm{{\rm k}})$ and that the angular momentum operator
in reciprocal k-space $\bm{{\mathcal{L}}}$ has the same form as in real
r-space (is Hermitian) and is given by $\bm{{\mathcal{L}}}=-i\bm{{\rm k}}
\times d/d\bm{{\rm k}}$,  i.e. $d/d\bm{{\rm k}}=
\bm{{\rm\hat{x}}}\partial_{k_x}+\bm{{\rm\hat{y}}}\partial_{k_y}+
\bm{{\rm\hat{z}}}\partial_{k_z}$ is the gradient operator in the Fourier
(reciprocal) space. Using this property and the Rayleigh
expansion $e^{i\bm{{\rm k}}'\cdot
\bm{{\rm r}}}=4\pi \sum_{l=0}^{\infty }i^lj_l(k'r)\sum_{m=-l}^{l}
Y_{lm} (\bm{{\rm\hat{r}}})Y_{lm}^*(\bm{{\rm\hat{k}}}')$ and making
$\bm{{\mathcal{M}}}_{lm} (\bm{{\rm k}})=j_l(kr)\bm{{\mathcal{X}}}_{lm}
(\bm{{\rm\hat{k}}})$, $\bm{{\mathcal{X}}}_{lm}(\bm{{\rm\hat{k}}})=
\bm{{\mathcal{L}}}Y_{lm}(\bm{{\rm\hat{k}}})/\sqrt{l(l+1)}$, we  obtain
\begin{eqnarray}
j_l(kr)
\left[
\begin{array}{c}
G_{lm}^{TE}\\
G_{lm}^{TM}
\end{array}
\right]
\!=\!\frac{i^l}{E_0 }\sqrt{\frac{2}{\pi}}
\int  d^3k'\bm{{\mathcal{M}}}_{lm}^*\cdot
\left[
\begin{array}{c}
\bm{\mathcal{E}}\\
Z\bm{\mathcal{H}}
\end{array}
\right]\!\!.
\end{eqnarray}

Now, only Fourier transforms of the form $k^2\bm{\mathcal{F}}(\bm{{\rm k}}')=
\delta (k'-k)\bm{\mathcal{F}}_{\rm k}(\bm{{\rm\hat{k}}}') $ will represent a
field $\bm{F}(\bm{{\rm r}})$ that satisfies the wave equation $\nabla^2\bm{F}+
k^2 \bm{F}=0$ in three dimensions. So we define the $\bm{{\rm\hat{k}}}$-only
dependent fields
\begin{eqnarray}
\left[
\begin{array}{c}
\bm{\mathcal{E}}(\bm{{\rm k}}')\\
\bm{\mathcal{H}}(\bm{{\rm k}}')
\end{array}
\right]
=\frac{\delta (k'-k)}{k'{}^2}\!
\left[
\begin{array}{c}
\bm{\mathcal{E}}_{\rm k}(\bm{{\rm\hat{k}}}')\\
\bm{\mathcal{H}}_{\rm k}(\bm{{\rm\hat{k}}}')
\end{array}
\right]\!\!.
\end{eqnarray}
Imposing this restriction one finally obtains that
\begin{eqnarray}\label{eq:newbsc}
\left[
\begin{array}{c}
G_{lm}^{TE}\\
G_{lm}^{TM}
\end{array}
\right]\!
=\!\frac{i^l}{E_0 }\sqrt{\frac{2}{\pi }}\int\!\!d\Omega_{k'}
\bm{\mathcal{X}}_{lm}^*(\bm{\hat{{\rm k}}}')\!\cdot\!
\left[
\begin{array}{c}
\bm{\mathcal{E}}_{\rm k}(\bm{{\rm\hat{k}}}')\\
Z\bm{\mathcal{H}}_{\rm k}(\bm{{\rm\hat{k}}}')
\end{array}
\right]\!\!.
\end{eqnarray}

To calculate the coefficients placed at arbitrary position $\bm{{\rm r}}_0$
independent of the coordinate system of the fields one can use the translation
 property of Fourier transform \cite{Arfken2005}.
Since this form is free of any radially-dependent function, our goal has been
achieved.

To calculate these Fourier transforms numerically it could be convenient to
use Laplace series \cite{Arfken2005}, also known as spherical harmonic
 transforms \cite{Suda2002}. Using the Laplace series expansion
$[\bm{\mathcal{E}}(\bm{{\rm\hat{k}}}'),Z\bm{\mathcal{H}}
(\bm{{\rm\hat{k}}}')]=\sum_{p',q'}[\bm{{\rm e}}_{p',q'},
\bm{{\rm h}}_{p',q'}]  Y_{p',q'}^*  (\bm{{\rm\hat{k}}}')$
one obtains
\begin{eqnarray}
\left[
\begin{array}{c}
G_{lm}^{TE}\\
G_{lm}^{TM}
\end{array}
\right]\!
=\!\frac{i^l}{E_0}\sqrt{\frac{2}{\pi }}\sum_{p',q'}\!
\left[
\begin{array}{c}
\bm{{\rm e}}_{p',q'}\\
\bm{{\rm h}}_{p',q'}
\end{array}
\right]
\!\cdot\!
\frac{{\left\langle p',q' \right|}\bm{{\mathcal{L}}}{\left|p,q\right\rangle}}
{\sqrt{l(l+1)}}.
\end{eqnarray}
The $\bm{{\rm e}}_{l',m'} /\bm{{\rm h}}_{l',m'} $ coefficients can be
calculated by several algorithms freely available in internet
and the  matrix ${\left\langle p',q'\right|}\bm{{\mathcal{L}}}
{\left|p,q\right\rangle} $ is  shown in most quantum mechanics books.

From now on we shall use the parameter $p=\pm(1)$ and will
write the components of the fiels in the complex circular
basis $(\bm{{\rm\hat{e}}}_-,\bm{{\rm\hat{z}}},\bm{{\rm\hat{e}}}_+)$, so
that the components are in the form $\bm{a}=[a_-,a_z,a_+]$ in which
$a_\pm=\bm{{\rm\hat{e}}}_\pm\cdot\bm{a}$ 
and $a_z=\bm{{\rm\hat{z}}}\cdot\bm{a}$. We also have that 
$\bm{{\rm\hat{e}}}_\pm=(\bm{{\rm\hat{x}}}\pm i\bm{{\rm\hat{y}}})/\sqrt{2}$.

\textbf{Plane wave: }
 The fields of a arbitrarily polarized plane wave are given by
$[\bm{{\rm E}}(\bm{{\rm r}}),Z\bm{{\rm H}}(\bm{{\rm r}})]=E_0
[\bm{\hat\epsilon},\bm{{\rm\hat{k}}}\times\bm{\hat\epsilon}]
e^{i\bm{{\rm k}}\cdot\bm{{\rm r}}}$, $\bm{\hat \epsilon }$
is the unit polarization vector, therefore the angular Fourier
transform is easily calculated, one obtains from equation
(\ref{eq:newbsc}) that
\begin{eqnarray}
\left[
\begin{array}{c}
G_{lm}^{TE}\\
G_{lm}^{TM}
\end{array}
\right]
=i^l 4\pi \bm{{\mathcal{X}}}_{lm}^*(\bm{{\rm\hat{k}}})\cdot
\left[
\begin{array}{c}
\bm{\hat \epsilon }\\
\bm{{\rm\hat{k}}}\times \bm{\hat \epsilon }
\end{array}
\right].
\end{eqnarray}
For the special case $\bm{{\rm\hat{k}}}=\bm{{\rm\hat{z}}}$ and for a circularly
polarized wave with 
$\bm{\hat \epsilon }=\bm{{\rm\hat{e}}}_p$ we have
\begin{eqnarray}\left[
\begin{array}{c}
G_{lm}^{TE }\\
G_{lm}^{TM }\end{array}\right]
=i^l\sqrt{2\pi \left(2l+1\right)}\delta_{lp}
\left[
\begin{array}{c}
1\\
\mp i
\end{array}
\right]\!\!,
\end{eqnarray}
$\delta_{n,m} $ is the Kronecker delta function. This is the result shown
 in Jackson's book \cite{Jackson1999} at less a $\sqrt{2}$ factor due to our
 choice of complex basis.

\textbf{General hollow waveguide mode: } The TM and TE modes of a hollow,
cylindrical waveguide of arbitrary cross-sectional shape are given as function
of $g(\bm{{\rm r}})=g_{{\rm \rho}}(\bm{{\rm \rho}})e^{ik_zz}$, 
$g_{{\rm \rho}}(\bm{{\rm \rho}})$ is the scalar solution of the transverse
wave equation   $(\partial_x^2+\partial_y^2+\gamma^2)g_{{\rm \rho}}
(\bm{{\rm \rho}})=0$ satisfying the boundary conditions at the waveguide
surfaces \cite{Jackson1999},
\begin{eqnarray}\label{eq:gwg1}
\left[
\begin{array}{c}
\bm{{\rm E}}^{TM}\\
Z\bm{{\rm H}}^{TE}
\end{array}
\right]
&=&\frac{k_z}{k}\bm{{\rm\hat{z}}}\times
\left[
\begin{array}{c}
-Z\bm{{\rm H}}^{TM}\\
\bm{{\rm E}}^{TE}\end{array}
\right]\\
\label{eq:gwg2}
&=&E_0 \left[\bm{{\rm\hat{z}}}+i\frac{k_z}{\gamma^{2}}\frac{d}{d\bm{\rho}}\right]
g(\bm{{\rm r}}),
\end{eqnarray}
 $k_z $ is the wavevector in the $z$
direction, $k^2=k_z^2+\gamma^2$, $\gamma $ is the transverse wavevector and 
$d/d\bm{\rho}=\bm{{\rm\hat{x}}}\partial_x+\bm{{\rm\hat{y}}}\partial_y$ 
is the transverse gradient operator.

It should be useful to introduce the cylindrical coordinate system in the
r-space and k-space. In the r-space we have $\bm{{\rm \rho}}=
x\bm{{\rm\hat{x}}}+y\bm{{\rm \hat{y}}}$, $\rho^2=\bm{{\rm \rho}}\cdot
\bm{{\rm \rho}}$ and $\bm{{\rm\hat{\rho}}}=\bm{{\rm\rho}}/\rho$. We have also
$\bm{{\rm \phi}}=-y\bm{{\rm \hat{x}}}+x\bm{{\rm \hat{y}}}$ and
$\bm{{\rm\hat{\phi}}}=\bm{{\rm\phi}}/\rho$. In the same way, in k-space we
change $\rho\to\gamma$ and $\phi\to\zeta$, and for spherical coordinates we 
also change $\theta\to\xi$ obtaining an equivalent system of
coordinates. So, the Fourier transforms of fields (\ref{eq:gwg1}) and
(\ref{eq:gwg2}) are given by
\begin{eqnarray}\label{eq:wgA}
\left[
\begin{array}{c}
\bm{\mathcal{E}}^{TM}\\
Z\bm{\mathcal{H}}^{TE}
\end{array}
\right]
&=&
{E_0 G(\bm{{\rm k}}')\left[\bm{{\rm\hat{z}}}-
\frac{k_z}{\gamma^2}\bm{\gamma }'\right]}\\
%
\label{eq:wgB}
\left[
\begin{array}{c}
\bm{\mathcal{E}}^{TE}\\
-Z\bm{\mathcal{H}}^{TM}
\end{array}
\right]
&=&E_0 \frac{k\gamma'}{\gamma^2}G(\bm{{\rm k}}')
\bm{{\rm\hat\zeta}}'.
\end{eqnarray}
Now, similar to the previous case, only $G_\rho(\bm{{\rm k}}')=
G_\gamma(\zeta')\delta(\gamma'-\gamma)/\gamma $ can represent in the Fourier
domain a scalar function $g_\rho(\bm{{\rm \rho}})$ that satisfies the
transverse wave equation $(\partial_x^2+\partial_y^2+\gamma^2)
g_\rho(\bm{{\rm \rho}})=0$. So, we have $G(\bm{{\rm k}}')=G_\gamma(\zeta')
\delta(\gamma'-\gamma)\delta(k_z'-k_z)/\gamma'=G_\gamma(\zeta')\delta(k'-k)
\delta(\zeta'-\zeta)/(\sin\xi'k'{}^2)$. Applying these results to
(\ref{eq:wgA}) and (\ref{eq:wgB}) and using
(\ref{eq:newbsc}) we obtain
\begin{eqnarray}\label{eq:bsc-wg1}
\left[
\begin{array}{c}
G_{lm}^{TE}[TM]\\
G_{lm}^{TM}[TE]
\end{array}\right]
&=&\frac{2i^lm}{\sqrt{l(l+1)}}\frac{k^2}{\gamma^2}Q_l^m
\left(\frac{k_z}{k}\right)G_\gamma^m\\
\label{eq:bsc-wg2}
\left[
\begin{array}{c}
G_{lm}^{TM}[TM]\\
-G_{lm}^{TE}[TE]
\end{array}
\right]
&=&
\frac{2i^{l-1}}{\sqrt{l(l+1)}}Q_l^m{}'\left(\frac{k_z}{k}\right)G_\gamma^m,
\end{eqnarray}
 $Q_l^m(x)$ is the normalized Associated Legendre 
 Polynomial in the way that $Y_l^m(\theta,\phi)=Q_l^m(\cos\theta)e^{im\phi}$
and $\cos\xi=k_z /k$, $Q_l^m{}'(x)$ is the derivative of $Q_l^m(x)$,
\begin{eqnarray}
\label{eq:gqgama}
G_\gamma^m&=&\int d\zeta'G_\gamma(\zeta')e^{-im\zeta'}.
\end{eqnarray}

\textbf{Rectangular hollow metallic waveguide mode: }
In this case there are two scalar functions $g_\rho$, given by
$g_\rho^{TE}(\bm{{\rm \rho}})=\cos(k_a^m x')\cos (k_b^ny')$ and
$g_\rho^{TM}(\bm{{\rm \rho}})=\sin (k_a^mx')\sin (k_b^n y')$,
 $k_a^m =m\pi/a$ and $k_b^n =n\pi/b$,
$\zeta^{mn}=\arctan(k_b^n/k_a^m)=\arctan(na/mb)$ and
$\gamma_{mn}=\pi\sqrt{m^2/a^2+n^2/b^2}$, $m$, $n$ integers and the origin at
the lower left corner of the waveguide \cite{Jackson1999}. If the origin is
placed at $(x_0,y_0,0)$ then $x'=x+x_0 $ and $y'=y+y_0$. The Fourier
transforms of these fields can be easily calculated, and the result is a
sum of four Dirac delta functions a the points $(k_a^m,k_b^n)$,
$(-k_a^m,k_b^n)$,
$(k_a^m,-k_b^n)$, $(-k_a^m,-k_b^n)$. These delta functions can be written in
cylindrical coordinates to give us the function $G_\gamma(\zeta')$. The
integral over $\zeta' $ given by $G_\gamma^q$ in (\ref{eq:gqgama}) -- 
the plus sign is for the TE-waveguide mode and the minus
sign is for the TM-waveguide mode -- is
\begin{eqnarray}\label{eq:gqg-rect}
G_\gamma^m=\alpha
\left(
e^{ im\zeta}\mathfrak{Re}\{i^me^{i\phi_-}\}\pm
e^{-im\zeta}\mathfrak{Re}\{i^me^{i\phi_+}\}\right)
\end{eqnarray}
in which $\alpha=\pi i^{-m} e^{ik_zz_0}$ and $\phi_\pm=k_xx_0\pm k_yy_0$.
The BSCs can now be found substituting (\ref{eq:gqg-rect}) in
(\ref{eq:bsc-wg1}) and (\ref{eq:bsc-wg2}).

\textbf{Cylindrical hollow metallic waveguide mode: }
The scalar solution for the electromagnetic fields in terms of cylindrical
coordinates with the origin on axis for the metallic waveguide are given by
$g_\rho(\bm{{\rm \rho}})=J_{m}(\gamma_{mn}\rho)e^{\pm im\phi}$,  $J_m(x)$
 are order $m$ Bessel functions, $J_m'(x)$ is the derivative of $J_m(x)$,
$\gamma_{m,n}=\chi_{m,n} /R$ or $\gamma_{m,n} =\chi'_{m,n} /R$, with \textit{R}
 being the radius of the cylinder, $\chi_{m,n} $ the $n$-th root of $J_{m} (x)$
 for the TM mode and $\chi'_{m,n} $ being the \textit{n}-th root of $J'_{m}(x)$
 for the TE mode \cite{Jackson1999}. From now on we define the functions
\begin{eqnarray}
\label{eq:psi}
\psi_m(\bm{{\rm k}};\bm{{\rm r}})&=&J_{m}(\gamma\rho)e^{ism\phi}e^{ik_zz}\\
\label{eq:Psi}
\Psi_m(\bm{{\rm k}};\bm{{\rm \hat{k}}}')&=&\sqrt{2\pi}(-i)^me^{ism\zeta'}
\frac{\delta(\xi'-\xi)}{\sin\xi'}
\end{eqnarray}
where again $s=\pm(1)$ and remember the addition theorem \cite{Abramowitz1970} 
to express the scalar function in terms of a new coordinate system
$\bm{{\rm r}}=\bm{{\rm r}}'+\bm{{\rm r}}_0$, written as a convolution
\begin{eqnarray}\label{eq:convol}
\psi_m(\bm{{\rm k}};\bm{{\rm r}}'+\bm{{\rm r}}_0)=\sum_{j=-\infty}^\infty\psi_{m-j}
(\bm{{\rm k}};\bm{{\rm r}}_0)\psi_j(\bm{{\rm k}};\bm{{\rm r}}').
\end{eqnarray}
The Fourier transform of $\psi_m(\bm{{\rm k}};\bm{{\rm r}})$
can be written as 
$\Psi_m(\bm{{\rm k}};\bm{{\rm \hat{k}}}')\delta(k'-k)/k'{}^2$. So, we can say
that for a cylindrical waveguide 
$g_\rho(\bm{\rho})=\psi_M(\bm{{\rm k}};\bm{{\rm r}})$, where $M$ denotes the
propagating mode and this way
\begin{eqnarray}
G_\gamma(\zeta')&=&\sqrt{2\pi}\sum_{j=-\infty}^\infty\psi_{M-j}(\bm{{\rm k}};
\bm{{\rm r}}_0)(-i)^je^{isj\zeta'}\\
\label{eq:gqg-cyl}
G_\gamma^m&=&2\pi(-i)^{sm}\psi_{M-sm}(\bm{{\rm k}};\bm{{\rm r}}_0).
\end{eqnarray}
The BSC's can now be found substituting (\ref{eq:gqg-cyl}) in
(\ref{eq:bsc-wg1}) and (\ref{eq:bsc-wg2}). The on-axis case can be obtained by setting
$\bm{{\rm r}}_0=0$, which implies that
$J_{M-sm}(\gamma_{mn}\rho_0)=\delta_{m,sM} $, and  therefore
\begin{eqnarray}\label{eq:gqg-cyl-oa}
G_\gamma^m=2\pi(-i)^{M}\delta_{m,sM}.
\end{eqnarray}
As we can see the sum over $m$ in (\ref{eq:expansion}) will disappear.

\textbf{Bessel Beams: }
We have calculated BSCs for two kind Bessel beams electromagnetic fields that
obey Maxwell equations \cite{McDonald1988}. They are derived using vector
potential and Lorentz gauge \cite{Davis1979} and are given as function of $\psi$
 already defined in (\ref{eq:psi}) and (\ref{eq:convol}) and $p=\pm(1)$.
\begin{eqnarray}
\left[
\begin{array}{c}
\bm{{\rm E}}_{z}\\
Z\bm{{\rm H}}_{z}
\end{array}
\right]
\!&=&\!E_0
\left[
\begin{array}{c}
\psi_M\bm{{\rm\hat{z}}}+
\nabla \nabla\cdot \left[\psi_M\bm{{\rm\hat{z}}}\right]/k^2\\
-i\nabla\times \left[\psi_M\bm{{\rm\hat{z}}}\right]/k
\end{array}
\right]\\
\left[
\begin{array}{c}
\bm{{\rm E}}_{p}\\
Z\bm{{\rm H}}_{p}
\end{array}
\right]
\!&=&\!E_0
\left[
\begin{array}{c}
\psi_{m-1}\bm{{\rm\hat{e}}}_p+\nabla\nabla\cdot
\left[\psi_{M-1}\bm{{\rm\hat{e}}}_p\right]/k^2\\
-i\nabla\times \left[\psi_{M-1}\bm{{\rm\hat{e}}}_p\right]/k
\end{array}
\right]\!\!.
\end{eqnarray}
The Fourier transform of these fields are given by
\begin{eqnarray}
\left[
\begin{array}{c}
\bm{\mathcal{E}}_{z}\\
Z\bm{\mathcal{H}}_{z}
\end{array}
\right]
&=&E_0\Psi_M
\left[
\begin{array}{c}
\bm{{\rm\hat{z}}}-\bm{{\rm \hat{k}}}'(\bm{{\rm \hat{k}}}'\cdot \bm{{\rm\hat{z}}})\\
\bm{{\rm \hat{k}}}'\times \bm{{\rm\hat{z}}}
\end{array}
\right]\\
\left[
\begin{array}{c}
\bm{\mathcal{E}}_{p}\\
Z\bm{\mathcal{H}}_{p}
\end{array}
\right]
&=&E_0\Psi_{M-1}
\left[
\begin{array}{c}
\bm{\hat e}_p-\bm{{\rm \hat{k}}}'(\bm{{\rm \hat{k}}}'\cdot\bm{\hat e}_p)\\
\bm{{\rm \hat{k}}}'\times \bm{\hat e}_p
\end{array}
\right]\!\!.
\end{eqnarray}
The $\bm{{\rm k}}'$ component is obviously null by orthogonality with
angular momentum. Using again the addition theorem (\ref{eq:convol})
and making $c_\pm^{lm}=\sqrt{l(l+1)-q(q\pm1)}$ one can show that the
coefficients are given by~(\ref{eq:bb1}) and (\ref{eq:bb2}).

In conclusion, we have shown that radially-independent amplitudes (the BSCs) of
a complete set of vector spherical wave functions  can be calculated explicitly
for an arbitrary electromagnetic field. We have shown how this result can be
used to determine the BSCs for several beam-types commonly employed in
photonics, although of course the method is not restricted to applications
within the field of optics. This new-found ability to evaluate the BSCs of the
VSWFs analytically makes it much easier to explore rapidly the influence of
experimental parameters in practical field scattering and optical tweezer
systems. With this analytical breakthrough, the long-standing problem of
evaluating the BSCs for an arbitrary field has been solved and the non-radial
dependence of the BSCs proven, allowing one to avoid unnecessary approximations
in the numerical evaluation of these quantities.
\begin{widetext}
\begin{eqnarray}\label{eq:bb1}
\left[
\begin{array}{c}
G_{lm}^{TE}\\
G_{lm}^{TM}
\end{array}
\right]_{z}
&=&\frac{4\pi i^{l-sm}}{\sqrt{l(l+1)}}\psi_{M-sm}(\bm{{\rm k}};\bm{{\rm r}}_0)
\left[
\begin{array}{c}
mQ_l^m(k_z/k)\\
i(\gamma/k)^2Q_l^m{}'(k_z/k)\\
\end{array}
\right]\\\label{eq:bb2}
\left[
\begin{array}{c}
G_{lm}^{TE}\\
G_{lm}^{TM}+ip(k_z/k)G_{lm}^{TE}
\end{array}
\right]_{p}
&=&\frac{4\pi i^{l-s(m-p)}}{\sqrt{2l(l+1)}}\psi_{M-1-s(m-p)}(\bm{{\rm k}};\bm{{\rm r}}_0)
\left[
\begin{array}{c}
c_{-p}^{lm}Q_l^{m-p}\\
ipm(\gamma/k)Q_l^m\\
\end{array}
\right]
\end{eqnarray}
\end{widetext}

\begin{thebibliography}{10}%
\makeatletter
\providecommand \@ifxundefined [1]{%
 \ifx #1\undefined \expandafter \@firstoftwo
 \else \expandafter \@secondoftwo
\fi
}%
\providecommand \@ifnum [1]{%
 \ifnum #1\expandafter \@firstoftwo
 \else \expandafter \@secondoftwo
\fi
}%
\providecommand \enquote [1]{``#1''}%
\providecommand \bibnamefont  [1]{#1}%
\providecommand \bibfnamefont [1]{#1}%
\providecommand \citenamefont [1]{#1}%
\providecommand\href[0]{\@sanitize\@href}%
\providecommand\@href[1]{\endgroup\@@startlink{#1}\endgroup\@@href}%
\providecommand\@@href[1]{#1\@@endlink}%
\providecommand \@sanitize [0]{\begingroup\catcode`\&12\catcode`\#12\relax}%
\@ifxundefined \pdfoutput {\@firstoftwo}{%
 \@ifnum{\z@=\pdfoutput}{\@firstoftwo}{\@secondoftwo}%
}{%
 \providecommand\@@startlink[1]{\leavevmode}%
 \providecommand\@@endlink[0]{}%
}{%
 \providecommand\@@startlink[1]{%
  \leavevmode
  \pdfstartlink
   attr{/Border[0 0 1 ]/H/I/C[0 1 1]}%
   user{/Subtype/Link/A<</Type/Action/S/URI/URI(#1)>>}%
  \relax
 }%
 \providecommand\@@endlink[0]{\pdfendlink}%
}%
\providecommand \url  [0]{\begingroup\@sanitize \@url }%
\providecommand \@url [1]{\endgroup\@href {#1}{\urlprefix}}%
\providecommand \urlprefix [0]{URL }%
\providecommand \Eprint[0]{\href }%
\@ifxundefined \urlstyle {%
  \providecommand \doi [1]{doi:\discretionary{}{}{}#1}%
}{%
  \providecommand \doi [0]{doi:\discretionary{}{}{}\begingroup
  \urlstyle{rm}\Url }%
}%
\providecommand \doibase [0]{http://dx.doi.org/}%
\providecommand \Doi[1]{\href{\doibase#1}}%
\providecommand \bibAnnote [3]{%
  \BibitemShut{#1}%
  \begin{quotation}\noindent
    \textsc{Key:}\ #2\\\textsc{Annotation:}\ #3%
  \end{quotation}%
}%
\providecommand \bibAnnoteFile [2]{%
  \IfFileExists{#2}{\bibAnnote {#1} {#2} {\input{#2}}}{}%
}%
\providecommand \typeout [0]{\immediate \write \m@ne }%
\providecommand \selectlanguage [0]{\@gobble}%
\providecommand \bibinfo [0]{\@secondoftwo}%
\providecommand \bibfield [0]{\@secondoftwo}%
\providecommand \translation [1]{[#1]}%
\providecommand \BibitemOpen[0]{}%
\providecommand \bibitemStop [0]{}%
\providecommand \bibitemNoStop [0]{.\EOS\space}%
\providecommand \EOS [0]{\spacefactor3000\relax}%
\providecommand \BibitemShut [1]{\csname bibitem#1\endcsname}%
\bibitem{Mie1908}%
  \BibitemOpen
  \bibfield{author}{%
  \bibinfo {author} {\bibfnamefont{G.}~\bibnamefont{Mie}},\ }%
  \bibfield{journal}{%
  \bibinfo {journal} {Ann. Phys.}\ }%
  \textbf{\bibinfo {volume} {330}},\ \bibinfo {pages} {377} (\bibinfo {year}
  {1908})%
  \bibAnnoteFile{NoStop}{Mie1908}%
\bibitem{Ashkin1970}%
  \BibitemOpen
  \bibfield{author}{%
  \bibinfo {author} {\bibfnamefont{A.}~\bibnamefont{Ashkin}},\ }%
  \bibfield{journal}{%
  \bibinfo {journal} {Phys. Rev. Lett.}\ }%
  \textbf{\bibinfo {volume} {24}},\ \bibinfo {pages} {156} (\bibinfo {year}
  {1970})%
  \bibAnnoteFile{NoStop}{Ashkin1970}%
\bibitem{Ashkin1987}%
  \BibitemOpen
  \bibfield{author}{%
  \bibinfo {author} {\bibfnamefont{A.}~\bibnamefont{Ashkin}}, \bibinfo {author}
  {\bibfnamefont{J.~M.}\ \bibnamefont{Dziedzic}},\ and\ \bibinfo {author}
  {\bibfnamefont{T.}~\bibnamefont{Yamane}},\ }%
  \bibfield{journal}{%
  \bibinfo {journal} {Nature}\ }%
  \textbf{\bibinfo {volume} {330}},\ \bibinfo {pages} {769} (\bibinfo {year}
  {1987})%
  \bibAnnoteFile{NoStop}{Ashkin1987}%
\bibitem{Collot1993}%
  \BibitemOpen
  \bibfield{author}{%
  \bibinfo {author} {\bibfnamefont{L.}~\bibnamefont{Collot}}, \bibinfo {author}
  {\bibfnamefont{V.}~\bibnamefont{Lefevreseguin}}, \bibinfo {author}
  {\bibfnamefont{M.}~\bibnamefont{Brune}}, \bibinfo {author}
  {\bibfnamefont{J.~M.}\ \bibnamefont{Raimond}},\ and\ \bibinfo {author}
  {\bibfnamefont{S.}~\bibnamefont{Haroche}},\ }%
  \bibfield{journal}{%
  \bibinfo {journal} {Europhys. Lett.}\ }%
  \textbf{\bibinfo {volume} {23}},\ \bibinfo {pages} {327} (\bibinfo {year}
  {1993})%
  \bibAnnoteFile{NoStop}{Collot1993}%
\bibitem{Vernooy1998}%
  \BibitemOpen
  \bibfield{author}{%
  \bibinfo {author} {\bibfnamefont{D.~W.}\ \bibnamefont{Vernooy}}, \bibinfo
  {author} {\bibfnamefont{A.}~\bibnamefont{Furusawa}}, \bibinfo {author}
  {\bibfnamefont{N.~P.}\ \bibnamefont{Georgiades}}, \bibinfo {author}
  {\bibfnamefont{V.~S.}\ \bibnamefont{Ilchenko}},\ and\ \bibinfo {author}
  {\bibfnamefont{H.~J.}\ \bibnamefont{Kimble}},\ }%
  \bibfield{journal}{%
  \bibinfo {journal} {Phys. Rev. A}\ }%
  \textbf{\bibinfo {volume} {57}},\ \bibinfo {pages} {R2293} (\bibinfo {year}
  {1998})%
  \bibAnnoteFile{NoStop}{Vernooy1998}%
\bibitem{Svoboda1994}%
  \BibitemOpen
  \bibfield{author}{%
  \bibinfo {author} {\bibfnamefont{K.}~\bibnamefont{Svoboda}}\ and\ \bibinfo
  {author} {\bibfnamefont{S.~M.}\ \bibnamefont{Block}},\ }%
  \bibfield{journal}{%
  \bibinfo {journal} {Ann. Rev. Biophys. Biomol. Struct.}\ }%
  \textbf{\bibinfo {volume} {23}},\ \bibinfo {pages} {247} (\bibinfo {year}
  {1994})%
  \bibAnnoteFile{NoStop}{Svoboda1994}%
\bibitem{Hell1993}%
  \BibitemOpen
  \bibfield{author}{%
  \bibinfo {author} {\bibfnamefont{S.}~\bibnamefont{Hell}}, \bibinfo {author}
  {\bibfnamefont{G.}~\bibnamefont{Reiner}}, \bibinfo {author}
  {\bibfnamefont{C.}~\bibnamefont{Cremer}},\ and\ \bibinfo {author}
  {\bibfnamefont{E.~H.~K.}\ \bibnamefont{Stelzer}},\ }%
  \bibfield{journal}{%
  \bibinfo {journal} {J. Microsc.}\ }%
  \textbf{\bibinfo {volume} {169}},\ \bibinfo {pages} {391} (\bibinfo {year}
  {1993})%
  \bibAnnoteFile{NoStop}{Hell1993}%
\bibitem{Brakenhoff1979}%
  \BibitemOpen
  \bibfield{author}{%
  \bibinfo {author} {\bibfnamefont{G.~J.}\ \bibnamefont{Brakenhoff}}, \bibinfo
  {author} {\bibfnamefont{P.}~\bibnamefont{Blom}},\ and\ \bibinfo {author}
  {\bibfnamefont{P.}~\bibnamefont{Barends}},\ }%
  \bibfield{journal}{%
  \bibinfo {journal} {J. Microsc.}\ }%
  \textbf{\bibinfo {volume} {117}},\ \bibinfo {pages} {219} (\bibinfo {year}
  {1979})%
  \bibAnnoteFile{NoStop}{Brakenhoff1979}%
\bibitem{Betzig1992}%
  \BibitemOpen
  \bibfield{author}{%
  \bibinfo {author} {\bibfnamefont{E.}~\bibnamefont{Betzig}}\ and\ \bibinfo
  {author} {\bibfnamefont{J.~K.}\ \bibnamefont{Trautman}},\ }%
  \bibfield{journal}{%
  \bibinfo {journal} {Science}\ }%
  \textbf{\bibinfo {volume} {257}},\ \bibinfo {pages} {189} (\bibinfo {year}
  {1992})%
  \bibAnnoteFile{NoStop}{Betzig1992}%
\bibitem{Sanchez1999}%
  \BibitemOpen
  \bibfield{author}{%
  \bibinfo {author} {\bibfnamefont{E.~J.}\ \bibnamefont{Sanchez}}, \bibinfo
  {author} {\bibfnamefont{L.}~\bibnamefont{Novotny}},\ and\ \bibinfo {author}
  {\bibfnamefont{X.~S.}\ \bibnamefont{Xie}},\ }%
  \bibfield{journal}{%
  \bibinfo {journal} {Phys. Rev. Lett.}\ }%
  \textbf{\bibinfo {volume} {82}},\ \bibinfo {pages} {4014} (\bibinfo {year}
  {1999})%
  \bibAnnoteFile{NoStop}{Sanchez1999}%
\bibitem{Cai2000}%
  \BibitemOpen
  \bibfield{author}{%
  \bibinfo {author} {\bibfnamefont{M.}~\bibnamefont{Cai}}, \bibinfo {author}
  {\bibfnamefont{O.}~\bibnamefont{Painter}},\ and\ \bibinfo {author}
  {\bibfnamefont{K.~J.}\ \bibnamefont{Vahala}},\ }%
  \bibfield{journal}{%
  \bibinfo {journal} {Phys. Rev. Lett.}\ }%
  \textbf{\bibinfo {volume} {85}},\ \bibinfo {pages} {74} (\bibinfo {year}
  {2000})%
  \bibAnnoteFile{NoStop}{Cai2000}%
\bibitem{Neves2006}%
  \BibitemOpen
  \bibfield{author}{%
  \bibinfo {author} {\bibfnamefont{A.~A.~R.}\ \bibnamefont{Neves}}, \bibinfo
  {author} {\bibfnamefont{A.}~\bibnamefont{Fontes}}, \bibinfo {author}
  {\bibfnamefont{L.~Y.}\ \bibnamefont{Pozzo}}, \bibinfo {author}
  {\bibfnamefont{A.~A.}\ \bibnamefont{de~Thomaz}}, \bibinfo {author}
  {\bibfnamefont{E.}~\bibnamefont{Chillce}}, \bibinfo {author}
  {\bibfnamefont{E.}~\bibnamefont{Rodriguez}}, \bibinfo {author}
  {\bibfnamefont{L.~C.}\ \bibnamefont{Barbosa}},\ and\ \bibinfo {author}
  {\bibfnamefont{C.~L.}\ \bibnamefont{Cesar}},\ }%
  \bibfield{journal}{%
  \bibinfo {journal} {Opt. Express}\ }%
  \textbf{\bibinfo {volume} {14}},\ \bibinfo {pages} {13101} (\bibinfo {year}
  {2006})%
  \bibAnnoteFile{NoStop}{Neves2006}%
\bibitem{Vollmer2002}%
  \BibitemOpen
  \bibfield{author}{%
  \bibinfo {author} {\bibfnamefont{F.}~\bibnamefont{Vollmer}}, \bibinfo
  {author} {\bibfnamefont{D.}~\bibnamefont{Braun}}, \bibinfo {author}
  {\bibfnamefont{A.}~\bibnamefont{Libchaber}}, \bibinfo {author}
  {\bibfnamefont{M.}~\bibnamefont{Khoshsima}}, \bibinfo {author}
  {\bibfnamefont{I.}~\bibnamefont{Teraoka}},\ and\ \bibinfo {author}
  {\bibfnamefont{S.}~\bibnamefont{Arnold}},\ }%
  \bibfield{journal}{%
  \bibinfo {journal} {Appl. Phys. Lett.}\ }%
  \textbf{\bibinfo {volume} {80}},\ \bibinfo {pages} {4057} (\bibinfo {year}
  {2002})%
  \bibAnnoteFile{NoStop}{Vollmer2002}%
\bibitem{Spillane2002}%
  \BibitemOpen
  \bibfield{author}{%
  \bibinfo {author} {\bibfnamefont{S.~M.}\ \bibnamefont{Spillane}}, \bibinfo
  {author} {\bibfnamefont{T.~J.}\ \bibnamefont{Kippenberg}},\ and\ \bibinfo
  {author} {\bibfnamefont{K.~J.}\ \bibnamefont{Vahala}},\ }%
  \bibfield{journal}{%
  \bibinfo {journal} {Nature}\ }%
  \textbf{\bibinfo {volume} {415}},\ \bibinfo {pages} {621} (\bibinfo {year}
  {2002})%
  \bibAnnoteFile{NoStop}{Spillane2002}%
\bibitem{Ng2005}%
  \BibitemOpen
  \bibfield{author}{%
  \bibinfo {author} {\bibfnamefont{J.}~\bibnamefont{Ng}}, \bibinfo {author}
  {\bibfnamefont{C.}~\bibnamefont{Chan}}, \bibinfo {author}
  {\bibfnamefont{P.}~\bibnamefont{Sheng}},\ and\ \bibinfo {author}
  {\bibfnamefont{Z.}~\bibnamefont{Lin}},\ }%
  \bibfield{journal}{%
  \bibinfo {journal} {Opt. Lett.}\ }%
  \textbf{\bibinfo {volume} {30}},\ \bibinfo {pages} {1956} (\bibinfo {year}
  {2005})%
  \bibAnnoteFile{NoStop}{Ng2005}%
\bibitem{Arlt2000}%
  \BibitemOpen
  \bibfield{author}{%
  \bibinfo {author} {\bibfnamefont{J.}~\bibnamefont{Arlt}}\ and\ \bibinfo
  {author} {\bibfnamefont{K.}~\bibnamefont{Dholakia}},\ }%
  \bibfield{journal}{%
  \bibinfo {journal} {Opt. Comm.}\ }%
  \textbf{\bibinfo {volume} {177}},\ \bibinfo {pages} {297} (\bibinfo {year}
  {2000})%
  \bibAnnoteFile{NoStop}{Arlt2000}%
\bibitem{Novotny1998}%
  \BibitemOpen
  \bibfield{author}{%
  \bibinfo {author} {\bibfnamefont{L.}~\bibnamefont{Novotny}}, \bibinfo
  {author} {\bibfnamefont{E.~J.}\ \bibnamefont{Sanchez}},\ and\ \bibinfo
  {author} {\bibfnamefont{X.~S.}\ \bibnamefont{Xie}},\ }%
  \bibfield{journal}{%
  \bibinfo {journal} {Ultramicroscopy}\ }%
  \textbf{\bibinfo {volume} {71}},\ \bibinfo {pages} {21} (\bibinfo {year}
  {1998})%
  \bibAnnoteFile{NoStop}{Novotny1998}%
\bibitem{Russell2003}%
  \BibitemOpen
  \bibfield{author}{%
  \bibinfo {author} {\bibfnamefont{P.~S.~J.}\ \bibnamefont{Russell}},\ }%
  \bibfield{journal}{%
  \bibinfo {journal} {Science}\ }%
  \textbf{\bibinfo {volume} {299}},\ \bibinfo {pages} {358} (\bibinfo {year}
  {2003})%
  \bibAnnoteFile{NoStop}{Russell2003}%
\bibitem{Euser2009}%
  \BibitemOpen
  \bibfield{author}{%
  \bibinfo {author} {\bibfnamefont{T.~G.}\ \bibnamefont{Euser}}, \bibinfo
  {author} {\bibfnamefont{M.~K.}\ \bibnamefont{Garbos}}, \bibinfo {author}
  {\bibfnamefont{J.~S.~Y.}\ \bibnamefont{Chen}},\ and\ \bibinfo {author}
  {\bibfnamefont{P.~S.~J.}\ \bibnamefont{Russell}},\ }%
  \bibfield{journal}{%
  \bibinfo {journal} {Opt. Lett.}\ }%
  \textbf{\bibinfo {volume} {34}},\ \bibinfo {pages} {3674} (\bibinfo {year}
  {2009})%
  \bibAnnoteFile{NoStop}{Euser2009}%
\bibitem{Gouesbet1988}%
  \BibitemOpen
  \bibfield{author}{%
  \bibinfo {author} {\bibfnamefont{G.}~\bibnamefont{Gouesbet}}, \bibinfo
  {author} {\bibfnamefont{B.}~\bibnamefont{Maheu}},\ and\ \bibinfo {author}
  {\bibfnamefont{G.}~\bibnamefont{Grehan}},\ }%
  \bibfield{journal}{%
  \bibinfo {journal} {JOSA-A}\ }%
  \textbf{\bibinfo {volume} {5}},\ \bibinfo {pages} {1427} (\bibinfo {year}
  {1988})%
  \bibAnnoteFile{NoStop}{Gouesbet1988}%
\bibitem{Gouesbet2009}%
  \BibitemOpen
  \bibfield{author}{%
  \bibinfo {author} {\bibfnamefont{G.}~\bibnamefont{Gouesbet}},\ }%
  \bibfield{journal}{%
  \bibinfo {journal} {J. Quant. Spectrosc. Radiat. Transfer}\ }%
  \textbf{\bibinfo {volume} {110}},\ \bibinfo {pages} {1223} (\bibinfo {year}
  {2009})%
  \bibAnnoteFile{NoStop}{Gouesbet2009}%
\bibitem{Nieminen2001}%
  \BibitemOpen
  \bibfield{author}{%
  \bibinfo {author} {\bibfnamefont{T.~A.}\ \bibnamefont{Nieminen}}, \bibinfo
  {author} {\bibfnamefont{H.}~\bibnamefont{Rubinsztein-Dunlop}},\ and\ \bibinfo
  {author} {\bibfnamefont{N.~R.}\ \bibnamefont{Heckenberg}},\ }%
  \bibfield{journal}{%
  \bibinfo {journal} {J. Quant. Spectrosc. Rad. Transfer}\ }%
  \textbf{\bibinfo {volume} {70}},\ \bibinfo {pages} {627} (\bibinfo {year}
  {2001})%
  \bibAnnoteFile{NoStop}{Nieminen2001}%
\bibitem{Mishchenko1996}%
  \BibitemOpen
  \bibfield{author}{%
  \bibinfo {author} {\bibfnamefont{M.~I.}\ \bibnamefont{Mishchenko}}, \bibinfo
  {author} {\bibfnamefont{L.~D.}\ \bibnamefont{Travis}},\ and\ \bibinfo
  {author} {\bibfnamefont{D.~W.}\ \bibnamefont{Mackowski}},\ }%
  \bibfield{journal}{%
  \bibinfo {journal} {J. Quant. Spectrosc. Rad. Transfer}\ }%
  \textbf{\bibinfo {volume} {55}},\ \bibinfo {pages} {535} (\bibinfo {year}
  {1996})%
  \bibAnnoteFile{NoStop}{Mishchenko1996}%
\bibitem{Neves2006a}%
  \BibitemOpen
  \bibfield{author}{%
  \bibinfo {author} {\bibfnamefont{A.~A.~R.}\ \bibnamefont{Neves}}, \bibinfo
  {author} {\bibfnamefont{A.}~\bibnamefont{Fontes}}, \bibinfo {author}
  {\bibfnamefont{L.~A.}\ \bibnamefont{Padilha}}, \bibinfo {author}
  {\bibfnamefont{E.}~\bibnamefont{Rodriguez}}, \bibinfo {author}
  {\bibfnamefont{C.~H.~B.}\ \bibnamefont{Cruz}}, \bibinfo {author}
  {\bibfnamefont{L.~C.}\ \bibnamefont{Barbosa}},\ and\ \bibinfo {author}
  {\bibfnamefont{C.~L.}\ \bibnamefont{Cesar}},\ }%
  \bibfield{journal}{%
  \bibinfo {journal} {Opt. Lett.}\ }%
  \textbf{\bibinfo {volume} {31}},\ \bibinfo {pages} {2477} (\bibinfo {year}
  {2006})%
  \bibAnnoteFile{NoStop}{Neves2006a}%
\bibitem{Jackson1999}%
  \BibitemOpen
  \bibfield{author}{%
  \bibinfo {author} {\bibfnamefont{J.~D.}\ \bibnamefont{Jackson}},\ }%
  \emph{\bibinfo {title} {Classical Electrodynamics}},\ \bibinfo {edition}
  {3rd}\ ed.\ (\bibinfo {publisher} {Wiley},\ \bibinfo {year} {1999})%
  \bibAnnoteFile{NoStop}{Jackson1999}%
\bibitem{Arfken2005}%
  \BibitemOpen
  \bibfield{author}{%
  \bibinfo {author} {\bibfnamefont{G.~B.}\ \bibnamefont{Arfken}}\ and\ \bibinfo
  {author} {\bibfnamefont{H.~J.}\ \bibnamefont{Weber}},\ }%
  \emph{\bibinfo {title} {Mathematical Methods for Physicists}},\ \bibinfo
  {edition} {international}\ ed.\ (\bibinfo {publisher} {Elsevier},\ \bibinfo
  {year} {2005})%
  \bibAnnoteFile{NoStop}{Arfken2005}%
\bibitem{Suda2002}%
  \BibitemOpen
  \bibfield{author}{%
  \bibinfo {author} {\bibfnamefont{R.}~\bibnamefont{Suda}}\ and\ \bibinfo
  {author} {\bibfnamefont{M.}~\bibnamefont{Takami}},\ }%
  \bibfield{journal}{%
  \bibinfo {journal} {Math. Comp.}\ }%
  \textbf{\bibinfo {volume} {71}},\ \bibinfo {pages} {703} (\bibinfo {year}
  {2002})%
  \bibAnnoteFile{NoStop}{Suda2002}%
\bibitem{Abramowitz1970}%
  \BibitemOpen
  \bibfield{author}{%
  \bibinfo {author} {\bibfnamefont{M.}~\bibnamefont{Abramowitz}}\ and\ \bibinfo
  {author} {\bibfnamefont{I.~A.}\ \bibnamefont{Stegun}},\ }%
  \emph{\bibinfo {title} {Handbook of mathematical functions}},\ \bibinfo
  {edition} {9th}\ ed.\ (\bibinfo {publisher} {Dover},\ \bibinfo {year}
  {1970})%
  \bibAnnoteFile{NoStop}{Abramowitz1970}%
\bibitem{McDonald1988}%
  \BibitemOpen
  \bibfield{author}{%
  \bibinfo {author} {\bibfnamefont{K.~T.}\ \bibnamefont{McDonald}}}%
   (\bibinfo {year} {1988}),\
  \Eprint{http://arxiv.org/abs/physics/0006046}{arXiv:physics/0006046v1}\
  \bibAnnoteFile{NoStop}{McDonald1988}%
\bibitem{Davis1979}%
  \BibitemOpen
  \bibfield{author}{%
  \bibinfo {author} {\bibfnamefont{L.~W.}~\bibnamefont{Davis}},\ }%
  \bibfield{journal}{%
  \bibinfo {journal} {Phys. Rev. A}\ }%
  \textbf{\bibinfo {volume} {19}},\ \bibinfo {pages} {1177} (\bibinfo {year}
  {1979})%
  \bibAnnoteFile{NoStop}{Davis1979}%
\end{thebibliography}
%

\end{document}